\documentclass[11pt]{article}
\usepackage{epsf}
\usepackage{epsfig}
\usepackage{graphicx}
\usepackage{amsmath}
\usepackage{amssymb}
\usepackage{color}
\usepackage{multicol}
\usepackage{psfrag}

\newcommand{\be}{\begin{equation}}
\newcommand{\ee}{\end{equation}}
\newcommand{\bea}{\begin{eqnarray}}
\newcommand{\eea}{\end{eqnarray}}
\newcommand{\beq}{\begin{equation}}
\newcommand{\eeq}{\end{equation}}
\newcommand{\ba}{\begin{array}}
\newcommand{\ea}{\end{array}}
\newcommand{\beqa}{\begin{eqnarray}}
\newcommand{\eeqa}{\end{eqnarray}}

\newcommand{\nn}{\nonumber}

\definecolor{Red}{rgb}{1.,0.,0.}

\begin{document}

\thispagestyle{empty}
\vskip 1.5 true cm

\renewcommand{\thefootnote}{\fnsymbol{footnote}}

\begin{center}
{\Large\bf $B \to D \tau \nu$ Branching Ratios:\\Opportunity for
Lattice QCD and Hadron Colliders}
  \\ [30 pt]
{\sc {\sc
J.~F.~Kamenik${}^{a,b}$
,
and F.~Mescia${}^{a}$
}}
 \\ [5pt]
{\sl ${}^a$INFN, Laboratori Nazionali di Frascati, Via E. Fermi
40, I-00044 Frascati, Italy} \\ [1 pt]
{\sl ${}^b$J. Stefan Institute, Jamova 39, P. O. Box 3000, 1001 Ljubljana, Slovenia}\\[35 pt]

{\bf Abstract} \\
\end{center}

\renewcommand{\thefootnote}{\arabic{footnote}}

\setcounter{footnote}{0}

\noindent In the Standard Model, scalar contributions to leptonic
and semileptonic decays are helicity suppressed. The hypothesis of
additional physical neutral/charged Higgses can enhance such scalar
contributions and give detectable effects especially in B physics.
 For the charged Higgs, experimental information on both
$Br(B \to D \tau \nu)$ and $Br(B\to\tau \nu)$ has already become
available and in particular the $B \to D \tau \nu$ branching ratio
measurements will be further improved in the coming years. Hadronic
uncertainties of  scalar contributions in semileptonic decays are
already in much better shape than the ones plaguing the helicity
suppressed leptonic decays $B\to\tau\nu$. Combining existing
experimental information form the B factories, we explore which
existing and future lattice estimates will be useful to directly
address new physics effects from  measurements of $Br(B_{u,d,s}\to
D_{u,d,s}\tau\nu)$, which can be performed also at hadron colliders.

\vskip 1.5 cm

As is often stressed, in the near future the LHC will represent the
main avenue to establish the presence of new physics by directly
detecting new particles at the TeV scale. On the other hand, virtual
effects of these particles can affect low-energy observables, probed
mainly by the flavor factories and soon by the LHCb. As has been
proven by the B-factories, the energy reach of such indirect
searches can often surpass direct detection strategies, making them
worthy of pursuit even at the opening of the new energy frontier.
Among the possible new particles, the Higgs boson is the only one
expected in the Standard Model (SM) picture. At the same time, we
have to observe that the established SM parametrization of the Higgs
sector is only a conservative example of a possible electro-weak
symmetry breaking mechanism. The present information on the massive
W and Z bosons from electro-weak precision tests only constrain the
goldstone modes~\cite{longhitano} of the Higgs field while leaving
space for an extended physical Higgs sector. Namely, additional
neutral/charged Higgses appear in many models trying to solve the
inconsistencies of the SM.

Therefore theoretical and experimental study of scalar effects in
observables, mediated at tree-level by neutral/charged
bosons\footnote{Loop induced flavor changing neutral current
processes, for example~\cite{haisch} $b\to s \gamma$, can be
sensitive to additional Higgses but this information is diluted by
contributions from other particles and final constraints are
model-dependent~\cite{paride-lunghi}.} is vitally important in
future experimental programmes. In particular, effective density
operators from charged scalar boson interactions have to be
considered in the effective weak Hamiltonian, which for $b\to q
(u,c)$ transitions, for example, reads \be \mathcal H_{eff}^{b\to q}
= \frac{G_F}{\sqrt 2} V_{qb} \sum_{\ell=
 e, \mu, \tau}\left[(\bar q \gamma_{\mu}(1-\gamma_5)b)\, (\bar \ell
\gamma^{\mu}(1-\gamma_5) \nu) + \,C^{\ell}_{NP} (\bar
q(1+\gamma_5)b) \,(\bar \ell(1-\gamma_5)\nu_\ell) \right]  +
\mathrm{h.c.}\,. \ee In the minimal flavor violating (MFV)
extensions of the SM~\cite{mfv} by an additional Higgs doublet the
additional new physics (NP) coupling can be written as \bea
C^{\ell}_{NP}&=&-\frac{m_b
m_{\ell}}{m^2_{H^{+}}}\frac{\tan^2\beta}{1+\epsilon_0\tan\beta}\,,
\label{eq:csp} \eea where $\tan\beta$ is the ratio of the two Higgs
VEVs while $\epsilon_0$ parameterizes possible Peccei-Quinn symmetry
breaking corrections and is typically of the order of $1\%$ in the
MFV minimal supersymmetric SM (MSSM). Due to the suppression of
quark and lepton Yukawa couplings in eq.~(\ref{eq:csp}), B
helicity-suppressed processes receive largest effects from the
charged Higgs. In this respect, the $B\to\tau\nu$ decay branching
ratio~\cite{gino}, given by \be\label{eq:btau} Br(B\to \tau\nu)
=\frac{G_F^2|V_{ub}|^2}{8 \pi} m^2_\tau f^2_B m_B
\left(1-\frac{m^2_\tau}{m_B^2}\right)^2 \times
\left\vert1+\frac{m_B^2}{m_b
m_{\tau}}\,C^{\tau}_{NP}\right\vert^2\,, \ee has often been stressed
as a good candidate and the recent B-factory results have given
important constraints on $C^{\tau}_{NP}$. Unfortunately, the
presently established experimental precision is only about $30\%$
and unlikely to improve in the near future as the perspectives to
measure $B\to\tau\nu$ at the Tevatron or LHCb are highly
compromised.  Furthermore, the SM expectation estimate presently
suffers from sizable parametrical uncertainties induced by
$|V_{ub}|$ and $f_B$. This opens the door for alternative modes to
be studied with the present experiments.

\begin{figure}[tb]
\begin{center}
\scalebox{1.1}{
\includegraphics{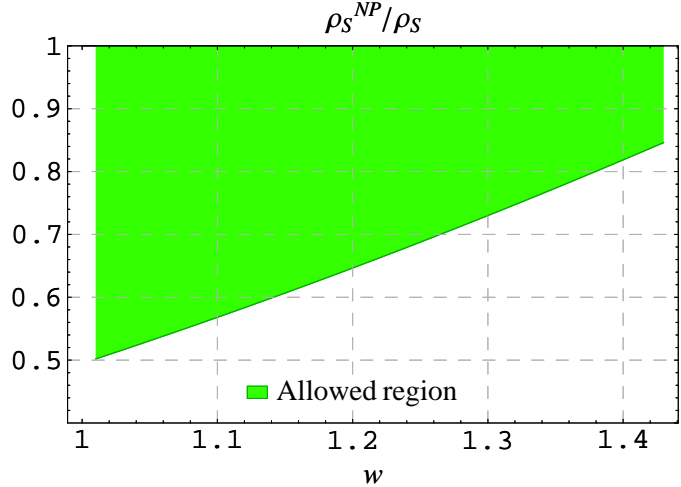}} \caption{\label{f0}
In green we plot as a function of $w$ the allowed region for
$\rho_S^{NP}(w)/\rho_S(w)$ in eq.~(\ref{eq:rho}), using constraints
from both $B\to \tau \nu$ and $K\to\mu\nu$
decays~\cite{ikado,flavia}. A large deviation from the unity, the SM
expectation, is still possible with respect to the SM. Note that
$\rho_{S}(w)$ contributes $50\%$ to the $Br(B\to D\tau\nu)$.}
\end{center}
\end{figure}

While Higgs effects in K
and D modes are small and difficult to disentangle at present
theoretical precision~\cite{flavia,stone}, the situation is much
better in the case of semileptonic $B\to D \ell \nu$
decays~\cite{bdtaunu,soni,trine}. The partial rate can be written in
terms of $w=v_B\cdot v_D$ as \bea\label{eq:width}
      \frac{d\Gamma(B\to D\ell\overline{\nu})}{dw}
        &=& \frac{G_F^2|V_{cb}|^2 m_B^5
        }{192\pi^3}\rho_V(w)\\
         &\times&\left[1 -
      \frac{m_{\ell}^2}{m_B^2}\,
      \left\vert1+
\frac{t(w)}{(m_b-m_c)m_{\ell}}\,C^{\ell}_{NP}\right\vert^2
     \rho_S(w) \right] ,\nn
\eea where $t(w) = m_B^2+ m_D^2 - 2w m_D m_B$ and we have decomposed
the rate into the vector and scalar Dalitz density contributions
\begin{eqnarray}
     \rho_V(w)  &=&
4\,\left(1+\frac{m_D}{m_B}\right)^2\left(\frac{m_D}{m_B}\right)^3\left(w^2-1\right)^{\frac{3}{2}}
         \left(1-\frac{m_\ell^2}{t(w)}\right)^2
           \left(1+\frac{m^2_\ell}{2 t(w)}\right)\,G(w)^2,\,\quad\quad\\
    \rho_S(w) & =& \frac{3}{2}
    \frac{m_B^2}{t(w)}
         \,
           \left(1+\frac{m^2_\ell}{2 t(w)}\right)^{-1}
         \frac{1+w}{1-w}\,\Delta(w)^2,
\end{eqnarray}
where $G(w)$ and $\Delta(w)$ encode our ignorance of the QCD
dynamics. Even before analyzing the theoretical uncertainties of
these modes let us note that the present constraints on
$C^{\tau}_{NP}$ from $K\to\mu\nu$~\cite{flavia} and
$B\to\tau\nu$~\cite{ikado} decays\footnote{In details, the
$\epsilon_0\tan\beta$ terms in eq.~(\ref{eq:csp}) are set to be
equal between $B\to \tau \nu$ and $K\to\mu\nu$, as it happens in MFV
MSSM.} still allow for sizable new physics effects in
eq.~(\ref{eq:width}) for the case of $B\to D\tau \nu$ as represented
in fig.~\ref{f0}, where the allowed region of the helicity
suppressed contribution of eq.~(\ref{eq:width}) for $B\to D\tau\nu$,
namely \be \label{eq:rho} \rho^{NP}_S(w)=\left\vert1+
\frac{t(w)}{(m_b-m_c)m_{\tau}}\,C^{\tau}_{NP}\right\vert^2
     \rho_S(w)\,,
     \ee
is shown.

\begin{figure}[tb]
\begin{center}
\scalebox{1.1}{\includegraphics{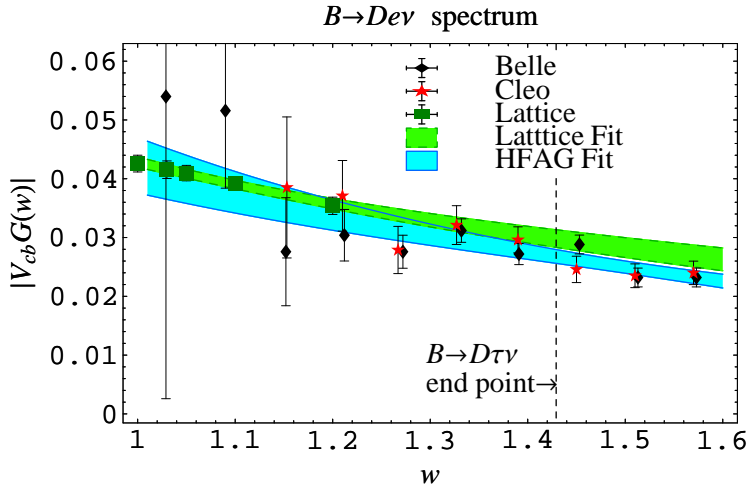}}
\caption{\label{fig:exp} Comparison of $B\to D e \nu$ form factor
determination from Belle~\cite{belle}, Cleo~\cite{cleo} and Lattice
QCD. The latter data points have been multiplied by the HFAG world
average value of $|V_{cb}|$ from inclusive measurements. The HFAG
average fit~\cite{hfag} to eq.~(\ref{eq:caprini}) is also shown.}
\end{center}
\end{figure}

The main parametric uncertainties in eq.~(\ref{eq:width}) are
represented by the modulus of $V_{cb}$ and the hadronic form factors
$G(w)$ and $\Delta(w)$. Presently, the most accurate value of
$|V_{cb}|=4.15(7)\%$ comes from the fit to inclusive $B\to X_c
\ell\nu$ decays which are insensitive to scalar
contributions~\cite{soni}. Because of charm states, information from
heavy quark expansion for the form factors is a priory
unsatisfactory, since corrections to the static limit $m_c, m_b\to
\infty$, formally parametrized by $\xi=1/m_b(1-m_b/m_c)$ can be
large and undetermined. More reliable information is expected from
the lattice and indeed a number of studies have computed the
normalization of the vector form factor $G(w)$ at $w=1$ to a
precision of a few percent, while a recent study extended its
determination to a region of $w\in [1,1.2]$~\cite{nazario}. These
values must then however be extrapolated over the entire
kinematically accessible decay phase space, which is larger in the
case of  $B\to D e \nu$ ($w\in [1,1.59]$) than for the tau mode
($w\in [1,1.43]$). For such an extrapolation, HFAG adopts the
parametrization of $G(w)$~\cite{Caprini:1997mu}~\footnote{Using
analyticity and crossing symmetry, a general parametrization for
semileptonic decays has been proposed in ref.~\cite{rhill}. However,
for modes such as $B\to D$, the smallness of z, and the judicious
use of heavy-quark symmetry in ref.~\cite{Caprini:1997mu}, allows
for a especially tailored parametrization in terms of
eq.~(\ref{eq:caprini})} \bea \label{eq:caprini} G(w) & = &
G(1)\times [1 - 8 \rho^2 z(w) + (51 \rho^2 -10)z(w)^2 - (252\rho^2 -
84)z(w)^3]\,,\eea with $z(w)=(\sqrt{w+1}-\sqrt 2)/(\sqrt{w+1}+\sqrt
2)$ in terms of two parameters: the normalization $G(1)$ and the
slope $\rho^2$. In addition, in the SM and as well as in its MFV
extensions only $G(w)$ will actually contribute to $B\to D e \nu$
and one can use experimental information on the differential decay
spectra in such an extrapolation\footnote{ For completeness, the
mechanism introduced in ref.~\cite{paride} to enhance electronic
modes in $K\to e \nu_\tau$ and $B\to e \nu_\tau$ by orders of
magnitude gives negligible effects less than $0.1\%$ for the partial
rate of $B\to D \ell \nu$, once the $K\to e \nu_\tau$
bound~\cite{flavia} is taken into account.}. At present, the
HFAG~\cite{hfag} experimental information consists of relatively old
publications by Belle~\cite{belle} and Cleo~\cite{cleo}. We can use
this information however to asses the relative precision obtainable
from combining lattice information with experimental inputs
efficiently. We compare in fig.~\ref{fig:exp}, the
Belle~\cite{belle} and Cleo~\cite{cleo} data on $|V_{cb} G(w)|$ and
the HFAG fit to the data from eq.~(\ref{eq:caprini}) (using
$|V_{cb}|G(1)=(42.3\pm 4.5)10^{-3}$, $\rho^2=1.17\pm 0.18$ with
correlation $0.93$), together with the lattice data from
ref.~\cite{nazario} and the fit from eq.~(\ref{eq:caprini}) to the
lattice results of $G(w)$~\cite{nazario} (yielding $G(1)=1.03(1)$,
$\rho^2=0.97(14)$) both multiplied by the HFAG value of $|V_{cb}|$
mentioned above.
The two sets are in agreement at present precision ($10\%$ on the
normalization and $15\%$ on the slope). Improvement however could
come from several sources: Babar has already announced to improve
the measurement of the differential decay rate to allow for
extraction of $\rho^2$ to below $10\%$ by reducing the statistics
error of Belle by a factor of 4~\cite{rotondo}\footnote{At this
level of precision, non-helicity suppressed NP contributions to the
$b\to c e \nu$ transition could be constrained for the first time
(for example R-parity violating MSSM~\cite{deandrea}).}. However, to
be able to apply this precision to the integrated rates, one would
need to precisely determine either $G(w)$ on the lattice while using
inclusive determination of $|V_{cb}|$ or consider ratios, where the
overall normalization factors of $|V_{cb}\,G(1)|$ cancel.

On the hand, the uncertainties coming from $\Delta(w)$, which
regulates the helicity suppressed terms, are already much smaller,
especially than those plaguing the dimensional variable $f_B$
entering $B\to \ell \nu$ decays. In other words, the current
(quenched) lattice estimate of $\Delta(w)$ for $w$ in the range
$1-1.2$ is at about $2\%$ precision, consistent with a constant
value of $\Delta(w)=0.46(1)$. Mainly, such an achievement on the
lattice was possible by introducing double ratios of lattice
correlators~\cite{hashimoto} and $\theta$ boundary
conditions~\cite{nazariobc}. Moreover, this precision can further be
improved by studies involving unquenched simulations and lighter sea
quark masses. In particular, a measurement of $B_s \to D_s \ell \nu$
will opt for lattice data on $B_s \to D_s$ form factors including
scalar contributions. These however no longer require chiral
extrapolations for the valence quarks, eliminating important sources
of systematics. Finally, since $\Delta(w)$ only contributes
significantly to the decays involving taus, the extrapolation from
the region presently probed by lattice simulations to the complete
kinematically accessible region is not large as is the case for the
$G(w)$ form factor in $B\to D e\nu$ transitions.

\begin{figure}
\begin{center}
\scalebox{0.4}{\includegraphics{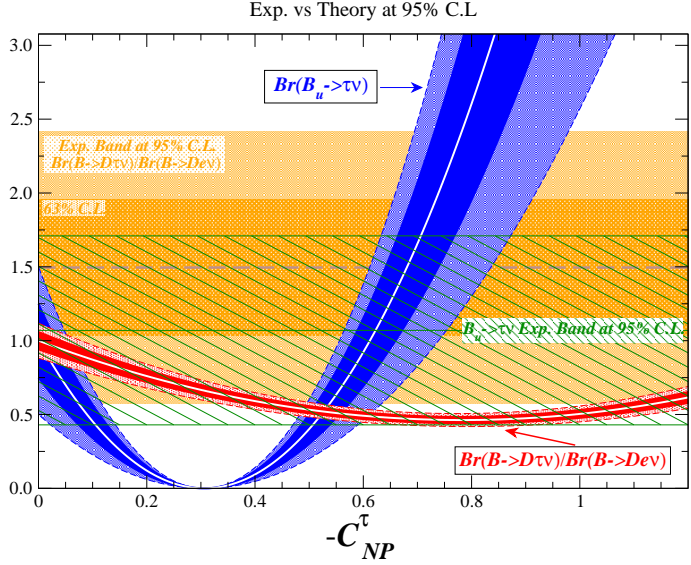}}
\caption{\label{fig:br-THDM} The ratio $Br(B \to D \tau \nu)/Br(B\to
D e \nu)$ is shown together with the $Br(B\to\tau\nu)$ as function
of $C^{\tau}_{NP}$, eq.~(\ref{eq:csp}). Both curves have been
normalized to their SM central values. Error bands on the curves
represent the theoretical uncertainties at $63\%$ and at $95\%$
C.L.. The horizontal bands represent the corresponding experimental
values~\cite{ikado,bexp}.}
\end{center}
\end{figure}

\begin{figure}
\begin{center}
\scalebox{0.75}{\includegraphics{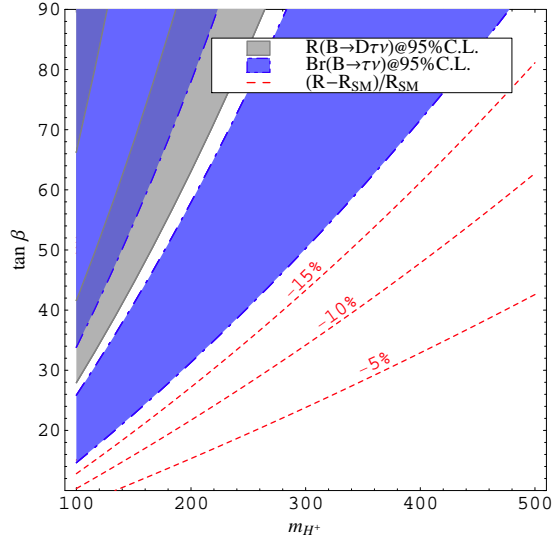}}
\caption{\label{fig:now}Exclusion region in the $m_{H^+} -
\tan\beta$ plane due to present determination of $B\to\tau\nu$ (in
blue) and $Br(B\to D\tau\nu)/Br(B\to D e\nu)$ (in gray). Note that
the small allowed band in the middle is excluded by $K\to \mu\nu$
determination~\cite{flavia} (not shown). Red dashed lines represent
percentage deviation from the SM prediction of $R=Br(B \to D \tau
\nu)/Br(B\to D e \nu)$ in the presently allowed region.}
\end{center}
\end{figure}

We finally combine these lessons and try to project the present
sensitivity of $B\to D\ell\nu$ decays to scalar contributions into
the near future. We start with the ratio $Br(B \to D \tau
\nu)/Br(B\to D e \nu)$~\cite{soni,trine} which, as stressed above,
even in the presence of NP scalar contributions only depends on two
hadronic parameters, the precision of which can furthermore be
improved in the near future: $\rho^2$ and $\Delta(w)$. By
integrating eq.~(\ref{eq:width}) with the use of
eq.~(\ref{eq:caprini}), the fitted lattice results for the form
factor, $\Delta(w)$ and the HFAG value of $\rho^2$ as determined
from the $B\to D e \nu$ spectrum, we average over the $B_{d,u}\to
D_{d,u}$ modes to obtain
\begin{eqnarray}
\frac{Br(B \to D \tau \nu)}{Br(B\to D e \nu)} &=& \left(0.28 \pm
0.02\right)\times \left[1 + 1.38(3) Re(C_{NP}^{\tau}) +
 0.88(2) |C_{NP}^{\tau}|^2\right]\,.
 \label{eq:th}
\end{eqnarray}
We see that the SM prediction uncertainty is already below $8\%$ and
is expected to be improved soon with the new Babar data on $\rho^2$.
Interestingly, Babar has already published a value~\cite{bexp} for
the above ratio with uncertainties of $30\%$, making it possible to
compare with the $B\to\tau\nu$ measurement and its bound on
$C^{\tau}_{NP}$ in fig.~\ref{fig:br-THDM}.
Even more importantly, unlike $B\to \tau \nu$, this measurement can
be improved at hadron colliders together with $B_s\to D_s \tau\nu$.
Therefore we plot the present exclusion region in the $\tan \beta$ -
$m_{H^+}$ plane in fig.~\ref{fig:now} together with the percentage
deviation from the SM prediction for $Br(B \to D \tau \nu)/Br(B\to D
e \nu)$ in the presently allowed region.

An even more prospective observable however, may be represented by
the ratio of partial $B\to D \tau (e) \nu$ decay widths integrated
over the same kinematical $w$ region. Since in the case of $B\to D
\tau \nu$ the kinematically available region is much smaller than
for the $B\to D e\nu$ one can just consider the full $Br(B\to D \tau
\nu)$~\cite{Itoh:2004ye}, while imposing a kinematical cut of
$w<1.43$ in the light lepton case. In this way one avoids the large
extrapolation away from the lattice data points and further reduces
the uncertainty due to the $\rho^2$ parameter. Presently such a
ratio can be estimated at $Br(B \to D \tau \nu)/Br(B\to D e
\nu)|_{w<1.43} = \left(0.56 \pm 0.02\right)\times \left[1 + 1.38(3)
Re(C_{NP}^{\tau}) +
 0.88(2) |C_{NP}^{\tau}|^2\right]$
with an error on the SM value of only $4\%$, while the relative new
physics contributions are not affected by the cut at all, since they
only appear in the tau mode. Once the experimental precision for
this observable would approach the above theoretical errors, one
could further restrict the kinematical region considered closer to
the one accessible to the lattice studies or finally consider binned
or differential rates.

\begin{figure}
\begin{center}
\scalebox{0.4}{\includegraphics{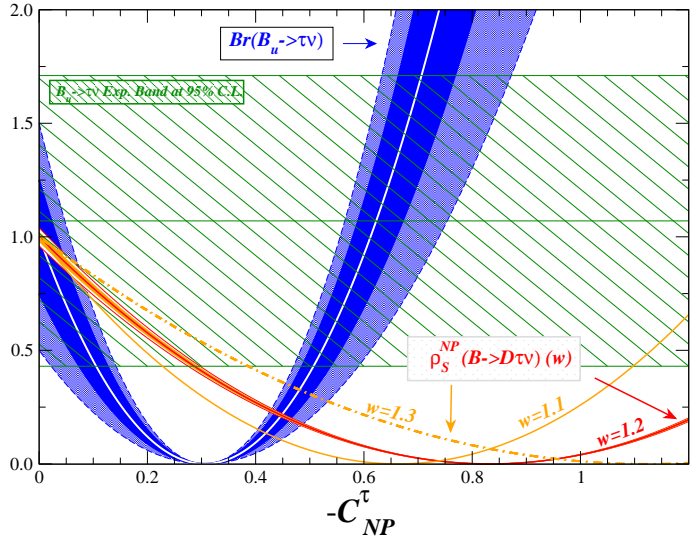}}
\caption{\label{fig:dg} The quantity $\rho^{NP}_S(w)$ from
eq.~(\ref{eq:rho}) is shown  for three values of $w$  as a function
of $C^{\tau}_{NP}$, eq.~(\ref{eq:csp}). The values of
$w=1.1,1.2,1.3$ are chosen to coincide with the presently available
lattice data~\cite{nazario}. Experimentally,
   $\rho^{NP}_S(w)$ can be accessed at a Super Flavour Factory
 via the measurement of
$d\Gamma(B \to D \tau \nu)/d\Gamma(B \to D \ell \nu)$,
eq.~(\ref{eq:width}) in those $w$-bins. }
\end{center}
\end{figure}

In existing literature, the differential rates~\cite{soni,trine} are
often stressed as being highly sensitive to scalar contributions in
$B\to D$ transitions compared to the integrated rate. However such
measurements will only become available with the advent of the Super
Flavor Factories, where both the $B\to D e \nu$ and $B \to D \tau
\nu$ spectra will be available at a few percent level in several $w$
bins. Then, measuring the ratio of $B \to D \tau \nu$ and $B\to D e
\nu$ differential distributions~\cite{soni} integrated over given
$w$-bins gives direct access to $\rho^{NP}_S(w)$ which can be
compared with the lattice estimates of $\rho_S(w)$ in the same bins
to obtain bounds on $C^{\tau}_{NP}$ by reducing ambiguities due to
$G(w)$ estimates and $w$ parameterizations. We project the
potentialities of measuring $\rho^{NP}_S(w)$ in eq.~(\ref{eq:rho})
with respect to $Br(B\to\tau\nu)$ in fig.~\ref{fig:dg}.

In the meantime, the ratio of (partialy) integrated rates $Br(B \to
D \tau \nu)/Br(B\to D e \nu)$ seems to represent the best strategy
for indirectly probing charged Higgs contributions to low energy
observables at the Tevatron and LHCb. Even if $Br(B \to D \tau
\nu)/Br(B\to D e \nu)|_{w<1.43}$ can not be measured directly,
precise data on $V_{cb}$ and $B\to D e\nu$ decay spectra from the B
factories can be used to obtain comparable precision directly on the
$B\to D\tau\nu$ branching ratio. Moreover, since the bounds from
$B\to\tau\nu$ are affected by larger theoretical uncertainties, the
$B\to D \tau\nu$ modes allow for an important crosscheck. Let's
mention that at $95\%$ with the present central value and with a
smaller experimental error of $20\%$, the exclusion region from
$B\to D \tau\nu$ is already competitive to the one from
$B\to\tau\nu$, while at $5\%$ error the SM and the MFV MSSM would
actually be excluded. Thus such a precise measurement of $Br(B\to D
\tau\nu)$ together with further lattice studies of $G(w)$ away from
$w=1$ and $\Delta(w)$ would be highly welcome since both the central
values as well as an accurate estimation of their errors are
essential to obtain valid bounds on new physics.

\section*{Acknowledgments}
We thank D.~Be\'cirevi\'c, G.~Isidori, M.~Rotondo, A.~Sarti and
A.~Annovi for discussions. This work is  supported in part by the EU
contract No.~MTRN-CT-2006-035482 (FlaviaNet).

\begin{multicols}{2}

\end{multicols}

\end{document}